\documentclass{emulateapj}

\shorttitle{Thermal emission} \shortauthors{Pe'er}

\newcommand{\beq}{\begin{equation}}
\newcommand{\eeq}{\end{equation}}
\newcommand{\ba}{\begin{array}}
\newcommand{\ea}{\end{array}}
\newcommand{\Li}{L_{52}}

\newcommand{\Gi}{\Gamma_{2}}

\newcommand{\D}{{\mathcal {D}}}

\newcommand{\Oo}{{\mathcal {O}}}
\def \etal{{\it et al.~}}

\begin{document}
\title{Temporal evolution of thermal emission from relativistically
  expanding plasma} 

\author{Asaf Pe'er\altaffilmark{1}\altaffilmark{2}}

\altaffiltext{1}{Space Telescope Science Institute, 3700 San Martin
  Dr., Baltimore, Md, 21218; apeer@stsci.edu}
\altaffiltext{2}{Giacconi Fellow}

\begin{abstract}

  Propagation of photons in relativistically expanding plasma
  outflows, ejected from a progenitor characterized by steady Lorentz
  factor $\Gamma$ is considered. Photons that are injected in regions
  of high optical depth are advected with the flow until they escape
  at the photosphere. Below the photosphere, the photons are coupled
  to the plasma via Compton scattering with the electrons. I show
  here, that as a result of the slight misalignment of the scattering
  electrons velocity vectors, the (local) comoving photon energy
  decreases with radius as $\varepsilon'(r) \propto r^{-2/3}$.  This
  mechanism dominates the photon cooling in scenarios of faster
  adiabatic cooling of the electrons.  I then show that the
  photospheric radius of a relativistically expanding plasma wind
  strongly depends on the angle to the line of sight, $\theta$. For
  $\theta \lesssim \Gamma^{-1}$, the photospheric radius $r_{ph}$ is
  $\theta$-independent, while for $\theta \gtrsim \Gamma^{-1}$,
  $r_{ph}(\theta) \propto \theta^2$. I show that the
  $\theta$-dependence of the photosphere implies that for flow
  parameters characterizing gamma-ray bursts (GRBs), thermal photons
  originating from below the photosphere can be observed up to tens of
  seconds following the inner engine activity decay. I calculate the
  probability density function $P(r,\theta)$ of a thermal photon to
  escape the plasma at radius $r$ and angle $\theta$. Using this
  function, I show that following the termination of the internal
  photon injection mechanism, the thermal flux decreases as
  $F_{BB}^{ob.}(t) \propto t^{-2}$, and that the decay of the photon
  energy with radius results in a power law decay of the observed
  temperature, $T^{ob.}(t) \propto t^{-\beta}$, with $\beta = 2/3$ at
  early times, which changes to $\beta \simeq 1/2$ later.  Detailed
  numerical results are in very good agreement with the analytical
  predictions. I discuss the consequences of this temporal behavior in
  view of the recent evidence for a thermal emission component
  observed during the prompt emission phase of gamma-ray bursts.

\end{abstract}


\section{Introduction}
\label{sec:intro}

Evidence for relativistic expansion in plasma winds exist in various
astronomical objects, such as microquasars \citep{MiRo94, HR95},
active galactic nuclei \citep[AGNs; ][]{LB85,GDSW06} and gamma-ray
bursts \citep[GRBs; ][]{Pac86,Good86}.  In many of these objects, the
density at the base of the flow is sufficiently high, so that the
optical depth to Thomson scattering by the baryon-related electrons
exceeds unity.  If the optical depth to scattering is high enough, the
emerging spectrum of photons emitted by radiative processes occurring
at or near the base of the flow is inevitably thermal or quasi-thermal
(a Wien spectrum could also emerge if the number of photons is
conserved by the radiative processes).  These photons escape the flow
once they decouple from the plasma, at the photosphere
\citep[e.g.,][]{Pac90}.

The photosphere is usually defined as a surface in space which
fulfills the following requirement: the optical depth to scattering a
photon originating from a point on this surface and reaching the
observer is equal to unity. Therefore, calculation of the position of
this surface requires knowledge of the density profile between this
surface and the observer (and, in principle, knowledge of the photon
energy and the velocity profile, since the cross section is energy
dependent; however, I will neglect these effects).  Calculation of the
photospheric surface in the case of steady, spherically
symmetric, relativistic wind was carried out by \citet{ANP91}. In this
work, it was found that the position of the photosphere has a
complicated, non-trivial shape which strongly depends on the viewing
angle and the wind Lorentz factor $\Gamma$. This shape can be
described analytically \citep[see][ eq. 3.4]{ANP91}.  For spherically
symmetric wind, the photospheric surface is symmetric with respect to
rotation around the axis to the line of sight. Thus, I will use the
term ``photospheric radius'' from here on to describe its position in
space, noting that the photospheric radius is a function of the angle
to the line of sight, $r_{ph}=r_{ph}(\theta)$. 

While the optical depth to scattering from the photospheric radius
$r_{ph}(\theta)$ to the observer is by definition
$\tau(r_{ph}[\theta]) = 1$, in fact photons have a finite probability
of being scattered at any point in space in which electrons
exist. Since in every scattering event a photon changes its
propagation direction and its energy, the observed flux and
temperature of the thermal photons depend on the last
scattering position, scattering time, the comoving temperature at this
position, and last scattering angle. A full description of the last
scattering position and scattering angle can only be done in terms of
probability density function $P(r,\theta)$. The probability density
function is an extension of the standard use of the photospheric
radius as a surface in space from which thermal photons emerge, to
consider the finite probability of a photon to emerge from an
arbitrary radius $r$ and arbitrary angle $\theta$.

Once the probability of a thermal photon to emerge at time $t$ from
radius $r$ and angle $\theta$ is known, the observed flux of the
thermal photons can be calculated.  The first observed (thermal)
photon originates from the radial axis towards the observer (on the
line of sight).  At later times an observer sees photons that
originate from increasingly higher angles to the line of sight and
from larger radii. The observed thermal flux thus varies with time.

The observed temperature of thermal photons emerging from radius $r$
at angle to the line of sight $\theta$ is blue shifted due to the
Doppler effect, $T^{ob.} = {\mathcal {D}} T'(r)$. Here, $T'(r)$ is the
photon temperature in the comoving frame, $T^{ob.}$ is the observed
temperature, ${\mathcal {D}} \equiv [\Gamma (1 -\beta \mu)]^{-1}$ is
the Doppler factor, $\beta c$ is the fluid velocity and $\mu =
\cos(\theta)$. Photons emitted on the line of sight are blue shifted
by $\D_0 = \D(\theta=0) \simeq 2 \Gamma$ (the last equality holds for
$\Gamma \gg 1$), while photons that originate from $\theta > 0$ are
blue shifted by Doppler factor $\D(\theta>0) < \D_0$. Note that
$\D(\theta)$ is a monotonically decreasing function of $\theta$.

The photon comoving temperature $T'(r)$ varies with the radius below
the photosphere. At $r<r_{ph}$, photons are coupled to the flow by
multiple Compton scattering. Therefore, once thermalize, the photons
comoving temperature is equal to the electrons comoving temperature
$T_{el}'(r)$ ($T_{el}'$ is measured in units of $m_e c^2$).  The
electrons comoving temperature changes as they propagate downstream,
due to adiabatic energy losses and possible internal heating
mechanisms that are coupled to the flow. The photon temperature is
thus expected to trace the electrons temperature. However, as I will
show below, an additional mechanism determines the photon temperature
below the photosphere. This mechanism is based on photon energy losses
due to the misalignment of the scattering electrons velocity vectors in
regions of high optical depth, and leads to photon comoving
temperature decay as a power law in radius, $T'(r) \propto r^{-2/3}$ 
(in the limit of relativistic outflows, characterized by Lorentz
factor $\Gamma \gg1$).  Therefore, as long as the electrons
comoving temperature does not drop faster than $T_{el}'(r) \propto
r^{-2/3}$, the photon comoving temperature traces the electrons
comoving temperature. However, if adiabatic energy losses cause the
electrons comoving temperature to decay faster than $r^{-2/3}$, than
the mechanism described below limits the photon comoving temperature
to decay as $r^{-2/3}$. In this case, the photon comoving temperature
does not follow the electrons temperature.

The mechanism by which photons lose their energy below the photosphere
is solely based on 2- and 3-dimensional scattering geometry.  In every
scattering event, the direction of the photon propagation vector
slightly changes, as the average scattering angle $\langle \theta
\rangle \sim \Gamma^{-1}$. Therefore, in every consequent scattering,
a photon is being scattered by electrons whos direction vectors are
slightly misaligned. This misalignment, in turn, inherently leads to
photon energy losses, regardless of the comoving temperature of the
electrons.  This effect can best be understood if one considers the
rest frame of the first scatterer.  Assuming that in this frame the
comoving photon energy is very low, $\varepsilon' \ll m_e c^2$, the
(comoving) outgoing photon energy is nearly equal to its energy before
the scattering, independent on the scattering angle. The misalignment
of the electrons velocity vectors implies that for scattering angles
which are different than $(0, \pi)$, the velocity vector of the next
scatterer points outward. Thus, the comoving energy of the photon in
the rest frame of the second scatterer is slightly lower than its
comoving energy in the rest frame of the first scatterer. As I will
show below, for constant outflow velocity characterized by Lorentz
factor $\Gamma \gg 1$, this effect results in a power law decay of the
comoving photon energy, $\epsilon'(r) \propto r^{-2/3}$. This decay,
in turn, results in a decrease of the observed temperature at late
times.

Transient physical sources have a finite emission duration, during
which their inner engine is active. Therefore, at any given observed
time $t^{ob.} > 0$, an observer sees simultaneously photons
originating from a range of radii and a range of angles to the line of
sight, $\theta_{\min} \leq \theta \leq \theta_{\max}$.  The dependence
of the photon comoving temperature on the photospheric (last scattering
event) radius and the dependence of both the photospheric radius and the
Doppler factor on the angle $\theta$ to the line of sight thus lead to
the conclusion that thermal emission (Planck spectrum) in the comoving
frame is observed as a modified black body.  However, the observed
spectrum is a convolution of black body spectra observed with
different fluxes and temperatures, and as such does not differ much
from a black body.

As long as the inner engine is active, the observed flux and
temperature are dominated by photons emitted on axis (on the line of
sight). Following the decay of the inner engine, the flux becomes
dominated by photons emitted from increasingly higher angles and radii
\citep[the curvature effect, also known as ``the high latitude
emission'' effect; see, e.g.,][for discussions on this effect in
optically thin emission models]{FMN96,KP00,RP02,Der04}. Considering
the optically thin cases, the results obtained in these works are not
valid for the scenario considered here, of thermal emission from {\it
  optically thick} expanding plasma.

The main goal of this paper is to provide a theoretical framework for
the analysis of thermal emission from astrophysical transient
characterized by relativistic outflows, and to calculate the expected
observed thermal flux and temperature at late times, following the
decay of the inner engine. A key motivation to this work are the
results obtained by \citet{Ryde04, Ryde05}, of a decaying thermal
component observed during the prompt emission phase of GRBs. In these
works, it was shown that after $\sim 1-3$~s the temperature of the
thermal component decreases as a power law in time, $T^{ob.} \propto
t^{-\alpha}$, with power law index $\alpha \simeq 0.6 - 1.1$. An
additional analysis (F. Ryde \& A.Pe'er 2008, in preparation) shows
that after a short rise, the flux of the black body component of these
bursts also decreases with time as $F_{BB}^{ob.} \propto t^{-\beta}$,
with power law index $\beta \approx 2.0 - 2.5$.  I show here that
these results are naturally obtained in a model which considers the
full spatial scattering positions and photon scattering angles (given
by the probability distribution function $P[r,\theta]$) and takes into
consideration the comoving energy losses of the photons below the
photosphere, due to the slight misalignment of the scattering
electrons.

This paper is organized as follows. I first calculate in
\S\ref{sec:r_ph} the optical depth at every point in space for
scattering into angle $\theta$, under the assumption of relativistic,
steady outflow. From this calculation, I find the angular dependence
of the photospheric radius, $r_{ph}(\theta)$. The calculation in this
section closely follows the treatment by \citet{ANP91}, although it is
somewhat more general since in this work the results were not
presented as a function of the viewing angle, $\theta$.  For
parameters characterizing GRBs, it is shown in \S\ref{sec:time} that
thermal emission can be observed up to tens of seconds.  The results
obtained in this section are used in \S\ref{sec:eps_prime} in which
the theory of photon energy loss due to misalignment of the scattering
direction vectors below the photosphere is developed.  The probability
density function $P(r,\theta)$ of a photon to be scattered from
radius $r$ and angle $\theta$ is introduced in
\S\ref{sec:analytic}. In this section, I calculate the expected flux
and temperature of the thermal emission, and show that following the
termination of the inner engine, the thermal flux and temperature decay as
power law in time, $F \propto t^{-\alpha}$, and $T^{ob.} \propto
t^{-\beta}$, with $\alpha=2$ and $\beta = 2/3$ at early times,
modified into $\beta \simeq 0.5$ later.  The numerical model is
presented in \S\ref{sec:numeric}. The Monte-Carlo simulation provides
full calculation of photon propagation in relativistically expanding
plasma, and is used to validate the approximations of the analytical
calculations.  The numerical results are compared to the analytical
predictions, and are found to be in very good agreement.  I summarize
the results and discuss their consequences in \S\ref{sec:summary}.

\section{Optical depth and photospheric radius in relativistically
  expanding plasma wind} 
\label{sec:r_ph}

Following the treatment of \citet{ANP91}, I consider the ejection of a
spherically symmetric plasma wind from a progenitor characterized by
constant mass loss rate $\dot M$, that expands with time independent
velocity $v = \beta c$. The ejection begins at $t=0$ from radius
$r=0$, thus at time $t$ the plasma outer edge is at radius $r_{out}(t)
= \beta c t$ from the center.  For constant ${\dot M}$ and $\Gamma$,
at $r<r_{out}$ the comoving plasma density is given by $n'(r) = {\dot
  M}/(4\pi m_p v \Gamma r^2)$, where $\Gamma = (1-\beta^2)^{-1/2}$. I
assume that emission of photons occurs deep inside the flow where the
optical depth $\tau \gg 1$ as a result of unspecified radiative
processes.  The emitted photons are coupled to the flow (e.g., via
Compton scattering), and are assumed to thermalize before escaping the
plasma once the optical depth becomes low enough.  In the following, I
will assume that the plasma wind occupies the entire space, i.e.,
$r_{out}(t) \rightarrow \infty$\footnote{The expansion of the plasma
  during the photon propagation implies that for photon emitted at
  $r_e < r_{out}(t)$, the plasma expands to radius $\sim \Gamma^2
  (r_{out}[t] - r_e)$, where $t$ is the photon emission time, before
  the photon crosses it. For parameters characterizing GRBs, the
  optical depth obtained by the full calculation converges to the
  result obtained using the approximation $r_{out} \rightarrow \infty$
  on an observed timescale of milliseconds. Full calculation implies
  that during this early expansion stage photons emitted from angle
  $\theta> \Gamma^{-1}$ are obscured; However, this calculation is
  omitted here, being relevant only for the very early stages of the
  expansion, and will appear elsewhere.}.

Calculation of the optical depth is done in the following way.
Consider a fluid element that moves in a particular direction (in the
observer frame) with constant velocity $\bf{v}$. I assume that the
cross section for photon scattering is energy independent, and is
equal to Thomson cross section, $\sigma_T$. This assumption holds as
long as in the (local) comoving frame of the fluid, the photon energy
is low, $\epsilon' \ll m_e c^2$. As a consequence of this assumption,
the mean free path of photons in the plasma comoving frame,
$l'=(n' \sigma_T)^{-1}$, is independent of the fluid velocity
$\bf{v}$, as measured in the observer frame.

Consider the propagation of photons through the medium in a direction
which makes an angle $\theta$ with respect to $\bf{v}$, in the
observer frame. The mean free path of photons as measured in this
frame is $l = l'/\Gamma(1-\beta \mu)$, where $\mu =
\cos(\theta)$. Fix two points along the light path, with distance
$ds$. If the fluid is at rest, $\beta=0$, the optical depth at
distance $ds$ is $d\tau_0 = n' \sigma_T ds$.  When the fluid velocity
is nonzero, the optical depth is
\beq 
d\tau = {ds \over l} = {l' \over l} d\tau_0 = \Gamma(1-\beta \mu) n'
\sigma_T ds.
\label{eq:dtau}  
\eeq

With equation \ref{eq:dtau} in hand, one can calculate the optical
depth for propagation of photons in spherically symmetric expanding
wind. Following the treatment of \citet{ANP91}, I define a cylindrical
coordinate system centered at the plasma expansion center, and assume
that the observer is located at plus infinity on the $z$ axis.
Consider a photon that propagates towards the observer (in the $+z$
direction) at distance $r_{\min}$ from the $z$ axis. Its distance from
the center will be denoted as $r$, $r=(r_{\min}^2 + z^2)^{1/2}$.  The
photon is assumed to be emitted at point $z_{\min}$ along the $z$
axis.  The optical depth measured along a ray traveling in the $+z$
direction and reaching the observer is given by
\beq
\ba{lcl}
\tau(r_{\min}, z_{\min}) & = & \int_{z_{\min}}^{\infty} n'(r) \sigma_T
\Gamma [1-\beta \cos(\theta)] dz \nonumber \\ 
& = &
{R_d \over \pi r_{\min}}{\bigg[}{\pi \over 2} - 
    \tan^{-1}\left({z_{\min} \over r_{\min}}\right) \nonumber \\
& & \qquad \quad -{\beta}\left. \left(1
    + {z_{\min}^2 \over r_{\min}^2}\right)^{-1/2}\right],
\ea
\label{eq:tau}
\eeq
where $\cos(\theta) = z/r = z/(r_{\min}^2+z^2)^{1/2}$, and equation
\ref{eq:dtau} was used. Here, 
\beq
R_d \equiv {{\dot M} \sigma_T \over 4 m_p \beta c}.
\label{eq:rd}
\eeq
Equation \ref{eq:tau} is identical to the result obtained by
\citet{ANP91}.

At the photon emission location, the angle 
$\theta(r_{\min},z=z_{\min})$ is the angle to the line of sight.
This allows to write equation \ref{eq:tau} in a simpler form. 
Using $\tan(\theta)=r_{\min}/z_{\min}$, one obtains
\beq
\tau(r,\theta) = {R_d \over \pi r}\left[{\theta \over \sin(\theta)} -
  \beta \right] \simeq {R_d \over 2 \pi r} \left( {1\over \Gamma^2}
+{\theta^2 \over 3}\right).
\label{eq:tau1}
\eeq
The last equality holds for $\Gamma \gg 1$ and small angle to the line
of sight, $\theta \ll \pi/2$, which allows the expansion 
 $\sin(\theta) \simeq \theta - \theta^3/6$.

\begin{figure}
\plotone{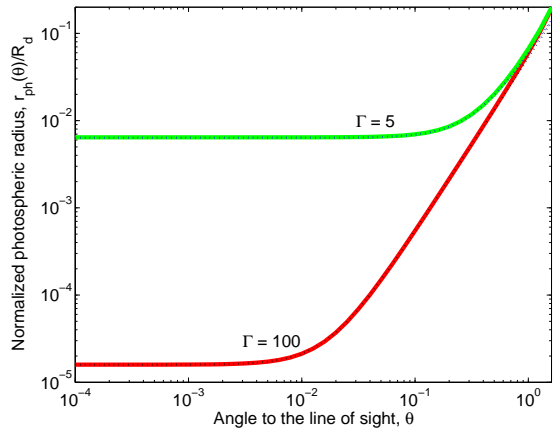}
\caption{Normalized photospheric radius $r_{ph}(\theta)/R_d$ as a
  function of the angle to the line of sight, $\theta$.  The thick
  lines show the exact solution of equation \ref{eq:tau1} for
  $\Gamma=100$ (lower line) and $\Gamma = 5$ (upper line). The dotted
  lines show the approximate solution $r_{ph}/R_d =
  (\theta^2/3+\Gamma^{-2})/2\pi$.  The approximate
  result in equation \ref{eq:r_ph} is nearly identical to the exact
  solution for $\theta \leq 1$.\bigskip }
\label{fig1}
\end{figure}

The photospheric radius is obtained by setting $\tau(r_{ph},\theta)
=1$,
\beq
r_{ph}(\theta) \simeq  {R_d \over 2 \pi} \left( {1\over \Gamma^2}
+{\theta^2 \over 3}\right).
\label{eq:r_ph}
\eeq
I thus find that at small viewing angle $\theta \ll \Gamma^{-1}$ the
photospheric radius is angle independent, $r_{ph} \simeq R_d/2\pi
\Gamma^2$, while for large angles $\theta \gg \Gamma^{-1}$, the
photospheric radius is $r_{ph}(\theta) \simeq R_d
\theta^2/6\pi$.\footnote{Note that often in the literature, one
  dimensional calculation is used for the photospheric radius. The
  result obtained, $R_d/2\pi \Gamma^2$ is correct in the limit
  $\theta \ll \Gamma^{-1}$.}  
The full calculation of the photospheric radius from
equation \ref{eq:tau1}, as well as the approximated solution in
equation \ref{eq:r_ph} are presented in figure \ref{fig1}. It is clear
from the figure that the approximate solution is nearly identical to
the exact solution in the range $\theta \lesssim 1$~radian.

\subsection{Characteristic time scale for observation of thermal
  emission in GRBs} 
\label{sec:time}

The strong angular dependence of the photospheric radius (equation
\ref{eq:r_ph}), implies that thermal photons originating from high
angles to the line of sight can be observed on a very long time scale
following the decay of the inner engine that produces the thermal
emission. Below the photosphere, the photons are coupled to the flow,
therefore their velocity component in the direction of the flow is
$\approx \beta c$.\footnote{One way to obtain this result is by noting
  that the average photon scattering angle (in the observer frame) is
  $\langle \theta \rangle \sim \Gamma^{-1}$. Thus, the photon velocity
  component in the flow direction is $c \times \cos \theta \simeq c
  (1- \theta^2/2) \sim \beta c$. This effect will be further discussed
  in \S\ref{sec:eps_prime} below.} Assuming that a photon is emitted
at $t=0$, $r=0$, it emerges from the photosphere at time
$t=r_{ph}(\theta)/\beta c$. Photons that propagate towards the observer
at angle to the line of sight $\theta$ are thus observed at a time
delay compared to a hypothetical photon that was emitted at $t=0$,
$r=0$ and did not suffer any time delay (``trigger'' photon), which is
given by $\Delta t^{ob.}(\theta) = [r_{ph}(\theta)/\beta c]\times [1 -
\beta \cos (\theta)]$.

For relativistic outflows, $\Gamma \gg1$, photons emitted on the line
of sight ($\theta=0$) are thus seen at a time delay with respect to the
trigger photon, 
\beq
\Delta t^{ob.}(\theta=0) \simeq {R_d \over 4 \pi \Gamma^4 \beta c}
\simeq 10^{-2} \Li \Gi^{-5} \, \rm{s}.
\label{eq:t_min}
\eeq
Here, ${\dot M} = L/\Gamma c^2$,  and
typical parameters characterizing GRBs, $L = 10^{52} \Li {\rm \, erg
  s^{-1}}$  and $\Gamma = 100 \Gi$ were used.

Photons emitted from high angles to the line of sight, $\theta \gg
\Gamma^{-1}$ (and $\theta \ll 1$) are observed at a much longer time
delay, 
\beq
t^{ob.}(\theta \gg \Gamma^{-1}) \simeq {R_d \over 3 \pi \beta
  c}\left({\theta^2 \over 2}\right)^2 \simeq 30 \, \Li \Gi^{-1}
\theta_{-1}^4 {\rm \, s}, 
\label{eq:t_max}
\eeq
where $\theta = 0.1 \theta_{-1}$.
The time scale derived on the right hand side of equation
\ref{eq:t_max} is based on the estimate of the jet opening angle in
GRB outflow, $\theta \leq \theta_j \simeq 0.1$  \citep[e.g.,][]{BKF03}.
One can thus conclude, that in relativistically expanding wind with
parameters characterizing emission from GRBs, thermal emission can be
observed up to tens of seconds following the decay of the inner engine.

\section{Photon energy loss due to the slight misalignment of the
  scattering electrons velocity vectors below the photosphere}
\label{sec:eps_prime}

Below the photosphere, photons undergo repeated Compton scattering
with the electrons in the flow. For a jet with finite,
constant opening angle, the velocity vectors of the electrons
propagating inside the jet are slightly misaligned. This, in turn,
leads to photon energy loss via repeated Compton scattering as the
photons propagate downstream. This mechanism is independent on the
adiabatic energy losses of the electrons, and thus dominates if the
electrons comoving temperature decreases faster than $r^{-2/3}$ (see
below).

In order to calculate the photon energy loss, I assume that the
comoving electrons temperature can be neglected, and take the limit
$T'_{el} = 0$. Consider a single scattering event between a photon and
an electron inside the flow. I explicitly assume that the photon
energy in the (local) comoving frame is low, $\varepsilon' \ll m_e
c^2$. In the scattering event, the photon is being scattered to angle
$\theta'$, in the (local) comoving frame. Denoting by $\varepsilon'$
the photon energy before the scattering (the incoming photon energy),
its outgoing energy is given by $\varepsilon_1' = \varepsilon' \left\{
  1 + (\varepsilon'/m_e c^2)\times [1-\cos(\theta')]\right\}^{-1}
\simeq \varepsilon'$. Thus, the photon {\it local} comoving energy is
not changed by a single scattering event\footnote{That is,
  Thomson limit is assumed for the scattering process.}.

\begin{figure}
\plotone{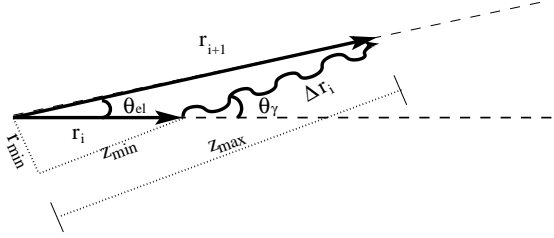}
\caption{Schematic view of two consequent scattering geometry in the lab
  (observer) frame. \vspace{12pt}
}
\label{fig:cartoon1}
\end{figure}

Consider two consequent scattering events, the first of which occurs
at radius $r_i$ and the second at radius $r_{i+1}$. Due to the
symmetry in the scattering direction, on the average the photon
propagation direction is parallel to the flow. Assuming that before
the first scattering the photon propagation direction is parallel to
the velocity vector of the first scatterer electron, following the
first scattering event, the photon propagation direction is at angle
$\theta_\gamma$ with respect to the first electron velocity vector (in
the lab frame), where $\cos(\theta_{\gamma}) = [\beta +
\cos(\theta')]/[1+\beta \cos(\theta')]$.  After being scattered, the
photon travels a distance $\Delta r_i$ until it is being scattered
again. The velocity vector of the consequent scatterer electron makes
an angle $\theta_{el}$ with the velocity vector of the first
electron (see figure \ref{fig:cartoon1}). The two electrons assume to
propagate at a similar Lorentz factor, $\Gamma$. As discussed above,
the photon comoving energy in the rest frame of the first scatterer,
$\varepsilon'_i$ is unchanged by the scattering process. However,
Lorentz transformation to the rest frame of the consequent scatterer
implies that the photon comoving energy in this frame is given by
\beq
\ba{lcl}
\varepsilon_{i+1}' & = & \varepsilon_i' {\Big\{} \Gamma^2\left[1 - \beta^2
      \cos(\theta_{el})\right] \nonumber \\
& & \quad \;  - \Gamma^2 \beta \left[
      \cos(\theta_{el})-1\right]{\left[ \cos(\theta_\gamma) - \beta
      \right] \over 1 - \beta \cos(\theta_\gamma) } \nonumber \\ 
 & & \quad \; \left.  - \Gamma \beta
    \sin(\theta_{el}) { \sin(\theta_\gamma) \over \Gamma\left[ 1 -
          \beta \cos(\theta_\gamma) \right]} \right\}.
\ea
\label{eq:eps2} 
\eeq
The photon scattering angle $\theta_{\gamma}$ and the angle between
the two electrons velocity vectors, $\theta_{el}$ are related via 
\beq
\tan(\theta_{el}) = { \Delta r_i \sin(\theta_\gamma) \over r_i + \Delta r_i
  \cos(\theta_\gamma)}
\label{eq:angels1}
\eeq
(see figure \ref{fig:cartoon1}).

Equation \ref{eq:eps2} can be considerably simplified by noting that
in the lab frame the photon scattering angle $\theta_\gamma \sim
\Gamma^{-1} \ll 1$, and that below the photosphere the average
distance traveled by the photon between consequent scattering is
small, $\Delta r_i/r_i \ll 1$ (see below).  Thus, one can approximate
$\tan(\theta_{el}) \approx \theta_{el} \approx (\Delta r_i /r_i)
\sin(\theta_\gamma)$ and $\cos(\theta_{el}) \approx 1 -
\theta_{el}^2/2$. With these approximations, equation \ref{eq:eps2}
becomes
\beq
\ba{lcl}
\varepsilon_{i+1}' & \simeq \varepsilon_i' & \Big\{  
1 +  {\Gamma^2\beta^2 \over 2} 
\left( {\Delta r_i \over r_i}\right)^2 \sin^2(\theta_\gamma) \nonumber \\
& &  \quad + 
{\Gamma^2 \beta \over 2} \left( {\Delta r_i \over r_i}\right)^2 {
  \sin^2(\theta_\gamma) \left[ \cos(\theta_\gamma) - \beta
      \right] \over 1 - \beta \cos(\theta_\gamma) } \nonumber \\
& & \quad \left. 
 - \beta \left({\Delta r_i \over r_i}\right) 
{ \sin^2(\theta_\gamma) \over 1
      -\beta \cos(\theta_\gamma)} \right\}.
\ea
\label{eq:eps3}
\eeq     

Of the three terms in the right hand side of equation \ref{eq:eps3}
that contribute to the energy change between consequent scatterings,
the last term is the dominant. Using $\sin(\theta_\gamma) \sim
\Gamma^{-1}= \Oo(\Gamma^{-1})$, one finds that $1-\beta
\cos(\theta_\gamma) = \Oo(\Gamma^{-2})$, and $\cos(\theta_\gamma)-
\beta = \Oo(\Gamma^{-2})$. Therefore, the first two terms are of the
order of $(\Delta r_i / r_i)^2$, while the last term is of the order
of $(\Delta r_i/r_i)$. Thus, for $(\Delta r_i/r_i) \ll 1$, equation
\ref{eq:eps3} is approximated as
\beq
\varepsilon_{i+1}' \approx \varepsilon_i' 
\left[  1 -\beta \left({\Delta r_i \over r_i}\right) 
{ \sin^2(\theta_\gamma) \over 1
      -\beta \cos(\theta_\gamma)} \right].
\label{eq:eps4}
\eeq

In order to determine the energy loss of a photon it is thus required
to estimate the ratio $(\Delta r_i/r_i)$. Calculation of this ratio is
done by transforming to the cylindrical coordinate system presented in
\S\ref{sec:r_ph}, in which the $z$ axis is the photon propagation
direction. The first of the two considered scattering occurs at
location $z_{\min}$ along this axis, and the proceeding scattering
occurs at location $z_{\max}$. Under these definitions, $\Delta r_i =
z_{\max} - z_{\min}$. The small scattering angle (in the lab frame)
implies that $\theta_{\gamma} \approx \tan(\theta_{\gamma}) =
r_{\min}/z_{\min}$, and that the radii of the consequent scatterings
can be approximated as $r_i \simeq z_{\min}$, $r_{i+1} \simeq
z_{\max}$ (see figure \ref{fig:cartoon1}).

Using these approximation in equation \ref{eq:tau1}, one finds that the
optical depth at the first scattering point is $\tau_i \approx (R_d/2
\pi z_{\min})\times [\Gamma^{-2} + (r_{\min}/z_{\min})^2/3]$, and at the
consequent scattering point it is $\tau_{i+1} \approx (R_d/2
\pi z_{\max})\times [\Gamma^{-2} + (r_{\min}/z_{\max})^2/3]$. 
Writing the optical depth between consequent scattering as $\Delta
\tau_i = \tau_i - \tau_{i+1}$ thus leads to  
\beq
\Delta \tau_i = {R_d \over 2 \pi \Gamma^2} 
\left( {1 \over z_{\min}} - {1 \over z_{\max}} \right) + 
r_{\min}^2 {R_d \over 6 \pi} \left( {1 \over z_{\min}^3} - {1 \over
    z_{\max}^3} \right).
\label{eq:dr1}
\eeq
Equation \ref{eq:dr1} can be simplified using $\Delta r_i = z_{\max} -
z_{\min}$ and the assumption $\Delta r_i \ll r_i$, 
\beq
\Delta \tau_i \simeq 
\Delta r_i {R_d \over 2 \pi z_{\min}^2} \left( {1\over \Gamma^2} +
  {r_{\min}^2 \over z_{\min}^2} \right) \approx
\Delta r_i {R_d \over 2 \pi r_i^2} 
\left( {1 \over \Gamma^2} + \theta_{\gamma}^2 \right).   
\label{eq:dr2}
\eeq

Since, on the average $\langle \Delta \tau_i \rangle = 1$, one obtains 
\beq
\langle \Delta r_i \rangle = 
{ r_i^2 \over {R_d \over 2 \pi} \left( {1 \over \Gamma^2} +
    \theta_{\gamma}^2 \right)}.
\label{eq:dr3}
\eeq
Comparison with equation \ref{eq:tau1} shows that equation
\ref{eq:dr3} justifies the assumption $\Delta r_i/r_i \ll 1$ used so
far, as long as the photon propagation occurs in region of high
optical depths. Using further this assumption in equations
\ref{eq:eps4} and \ref{eq:dr3}, the (local) photon comoving energy and
radius after $n$ scattering events can be approximated as
\beq
\ba {lcl}
\varepsilon_n' & \approx & \varepsilon_1' e^{-n r_1 \langle \omega
  \rangle}; 
\nonumber \\
r_n & \approx & r_1 e^{n r_1 \langle \alpha \rangle}.
\ea
\label{eq:exp}
\eeq
Here, $\varepsilon_1'$, $r_1$ are the photon comoving energy and
radius at the point in which the photon is introduced into the plasma,
$ \langle \omega \rangle$ and $\langle \alpha
\rangle$ are the average of the functions 
\beq
\ba{lcl}
\langle \omega \rangle & = & 
{\bigg \langle} {\beta \sin^2(\theta_{\gamma}) \over {R_d \over 2 \pi}\left(
    {1\over \Gamma^2} + \theta_{\gamma}^2 \right)[1-\beta
    \cos(\theta_{\gamma})]} {\bigg \rangle} ; \smallskip \nonumber \\
\langle \alpha \rangle & = & 
{\bigg \langle} {1 \over {R_d \over 2 \pi} \left( {1 \over \Gamma^2}
    + \theta_{\gamma}^2 \right) } {\bigg \rangle}
\ea
\label{eq:indx}
\eeq
over the scattering events angles. Note that both 
$\omega$ and $\alpha$ are functions only of the photon scattering 
angle, $\theta_{\gamma}$ (for constant flow parameters, $R_d$ and
$\Gamma$, as assumed here). 

Equation \ref{eq:exp} implies that the comoving energy of photons
decreases with radius as  
\beq
\varepsilon' (r) \propto r^{-{\langle \omega \rangle \over \langle
    \alpha \rangle}}.
\label{eq:eps_r}
\eeq
The values of the functions $ \langle \omega \rangle$ and $\langle \alpha
\rangle$  are calculated in \S\ref{sec:flux_temp} below.

\section{Late time decay of the observed temperature and flux of the
  thermal emission}
\label{sec:analytic}

As long as the radiative processes that produce the thermal photons
deep inside the flow are active, the observed thermal radiation is
dominated by photons emitted on the line of sight towards the
observer. Once these radiative processes are terminated, the radiation
becomes dominated by photons emitted off axis and from larger radii on
a very short time scale (see eq. \ref{eq:t_min}).  In order to
calculate the observed thermal flux and temperature at late times, it
is thus required to calculate the probability of a photon to be
emitted from radius $r$ and angle to the line of sight $\theta$.
Since thermal photons are coupled to the flow below the photosphere,
the emission radius of these photons is in fact the radius in which
the last scattering event takes place.

In the calculation below, I assume that photons are coupled to the
flow via Compton scattering below the last scattering event radius,
$r$. Below this radius, the photons velocity vector is, on the
average, parallel to the direction of the flow. At the last scattering
event, the photon is being scattered into angle $\theta$ with respect
to the direction of the flow. Since below $r$ the velocity component
of the photon in the direction of the flow is $\approx \beta c$, a
photon emitted at time $t=0$ decouples from the plasma at time
$t_{dcp} \simeq r/\beta c$. This photon is observed at a delay with
respect to the ``trigger'' photon that was emitted at $t=0$, $r=0$ and
propagates towards the observer by $t^{ob.} \equiv \Delta t^{ob.} =
r/{\beta c} \times [1 - \beta \cos(\theta)] \equiv r u /{\beta c} $.
Denoting by $T'(r)$ the comoving temperature of the photons at radius
$r$, the observed temperature of the photons is $T^{ob.} = T'(r) \D$,
where $\D = [\Gamma(1-\beta \mu)]^{-1}$ is the Doppler factor, and
$\mu \equiv \cos(\theta)$. The observed flux and temperature are thus
functions of $r$ and $\theta$ (or $u$). 

The calculation of the photospheric radius presented in
\S\ref{sec:r_ph} gives, by definition, the radius above which the
optical depth to scattering is equal to unity. Photons, however, have
a finite probability of being scattered at any point in space in which
electrons exist. Thus, the calculation presented in \S\ref{sec:r_ph}
needs to be extended to include the finite probability of photons to
be emitted at any radius and into arbitrary angle. I introduce here
calculation of this probability density function $P(r,\theta)$ under
some simplified approximations, which allow full analytic calculation
of the flux and temperature at late times. The results of an exact
numerical calculation are presented in \S\ref{sec:numeric} below. It
is shown there that the analytical calculations presented here are in
very good agreement with the exact numerical results.

\subsection{Probability of photons to decouple from the plasma at
  radius $r$ and be scattered into angle $\theta$}
\label{sec:P_r_t}

The increase of the photospheric radius with the angle to the line of
sight, $\theta$ (see eq. \ref{eq:r_ph}) implies that the probability
of a photon to be scattered at angle $\theta$ is
$r$-dependent. Nonetheless, as suggested by the numerical results in
\S\ref{sec:numeric} below, this dependence is limited to a cutoff at a
maximum angle from which photons are observed $ \theta_{\max}(r)$, and
does not affect much the probability of photons to be scattered to
smaller angles, $\theta < \theta_{\max}$. In the model below, I thus
make a separation of variables to write $P(r,\theta) = P(r) \times
P(\theta)$. The validity of this assumption (as well as the other
approximations used) is verified by the numerical results presented in
\S\ref{sec:numeric}.

The probability of the last scattering event of a thermal photon to
occur at radius $r.. r+ \delta r$ is calculated in the following way.
The optical depth $\tau(r,\theta)$ for a photon scattered at radius
$r$ into angle $\theta$ to reach the observer was calculated in
equation \ref{eq:tau1}. In determining the probability of a photon to
be scattered from radius $r.. r+ \delta r$, the angle into which the
scattering occurs is of no importance \footnote{Due to symmetry, the
  probability of a photon propagating on the line of sight to be
  scattered into angle $\theta$, is equal to the probability of a
  photon propagating at angle $\theta$ with the line of sight to be
  scattered into the line of sight.} . One can therefore write the
dependence of the optical depth on the radius as $\tau(r) \propto
r^{-1}$ (see eq. \ref{eq:tau1}). This optical depth is the integral
over the scattering probability of a photon propagating from radius
$r$ to $+\infty$, i.e., $\tau(r) = \int_r^{\infty} (d\tau/dr) dr$,
from which it is readily found that $(d\tau/dr)|_r \propto r^{-2}$.
As the photon propagates from radius $r$ to $r + \delta r$, the
optical depth in the plasma changes by $\delta \tau = (d\tau/dr)|_r
\delta r$. Therefore, the probability of a photon to be scattered as
it propagates from radius $r$ to $r + \delta r$ is given by
\beq
P_{sc.} ( r .. r + \delta r) = 1 - e^{-\delta \tau} \approx \delta
\tau \propto {\delta r \over r^2}.
\label{eq:P_r1} 
\eeq    

For the last scattering event to take place at $r.. r+ \delta r$, it
is required that the photon will not undergo any additional scattering
before it reaches the observer. The probability that no additional
scattering occurs from radius $r$ to the observer is given by
$\exp(-\tau[r])$. The probability density function $P(r)$ for the last
scattering event to occur at radius $r$, is therefore written as 
\beq
P(r) = {r_0 \over r^2} e^{-(r_0/r)}.
\label{eq:P_r}
\eeq 
The function $P(r)$ in equation \ref{eq:P_r} is normalized,
$\int_0^{\infty} P(r) dr = 1$.  By comparison to equation
\ref{eq:tau1}, the proportionality constant is $r_0 \equiv
r_{ph}(\theta=0) = R_d/2\pi\Gamma^2$.

The probability of a photon to be scattered into angle $\theta$ is
calculated as follows.  Since photons undergo multiple Compton
scattering below the photosphere, I assume that before the last
scattering event the photon propagation direction is parallel to the
flow. In the following, I neglect the dipole approximation in the last
scattering event angle, $\theta'$.  I thus assume that in the local
comoving frame the scattering is isotropic, i.e., $d\sigma/d\Omega' =
Const$. This approximation was checked numerically to be valid (see
\S\ref{sec:numeric} below).  Since the spatial angle $d\Omega' = \sin
\theta' d\theta' d \phi'$, the probability of a photon to be scattered
to angle $\theta'$ (in the comoving frame) is $dP/d\theta' \propto
\sin \theta'$. Integrating over the range $0 \leq \theta' \leq \pi$
gives the normalization factor $1/2$. Thus, the isotropic scattering
approximation leads to $P(\theta') = (\sin \theta')/2$.

Using Lorentz transformation to the observer frame, the probability of
scattering into angle $\theta$ with respect to the flow direction,
which is the observed angle to the line of sight is
\beq
\ba{lcl}
P(\theta) & = & P(\theta') {d\theta' \over d\cos\theta'} 
{d  \cos\theta' \over d \cos\theta} {d \cos\theta \over d\theta} \smallskip \nonumber \\
& = &
{\sin\theta \over 2 \Gamma^2 (1-\beta \cos\theta)^2}. 
\ea
\label{eq:P_theta}
\eeq
Using the definition $u \equiv 1-\beta \cos \theta $, equation
\ref{eq:P_theta} becomes
\beq
P(u) = { 1 \over 2 \Gamma^2 \beta u^2}.
\label{eq:P_u}
\eeq 
Note that $1-\beta \leq u \leq 1+\beta$, and the function $P(u)$
in equation \ref{eq:P_u} is normalized, $\int_{1-\beta}^{1+\beta} P(u)
du = 1$.

\subsection{Temporal evolution of the observed flux and temperature at
late times}
\label{sec:flux_temp}

The diffusion model presented above implies that thermal photons
emerging from the expanding plasma (i.e., last scattered) at radius
$r$ and into angle $\theta$ are observed at time $t^{ob.} = r u /
\beta c$. This assumption implies that all the photons are introduced
into the plasma (by radiative processes occurring deep inside the
flow) at the same instance, i.e., the photon injection function is
assumed to be a $\delta$-function in time. For finite photon injection
function, the result is a convolution of the $\delta$-function
calculation presented here. For a quick termination of the inner
engine, the $\delta$-function approximation leads to a good
description of the temporal evolution of temperature and flux at late
times, once the inner engine terminates.

Following the decay of the inner engine, the observed
flux at time $t^{ob.}$ is proportional to the probability of photon
emission from radius $r$ and into angle $\theta$ by
\beq 
\ba{lcl}
F(t^{ob.}) & = & F_0 \int_{r_{\min}}^{r_{\max}} P(r) dr 
\int_{u_{\min}}^{u_{\max}}
P(u) du \times \delta \left( t^{ob.} = {r u \over \beta c} \right) \nonumber
\\
& = & F_0  \int_{r_{\min}}^{r_{\max}} {r_0 \over r^2} e^{-(r_0/r)} dr  
\int_{1-\beta}^{1+\beta} {1 \over 2 \Gamma^2 \beta u^2} du \nonumber \\
& &  \times
   \left({\beta c \over r}\right) 
 \delta \left( u = {\beta c t^{ob.} \over r} \right) \nonumber \\
& = & F_0 {r_0 \over 2\Gamma^2 \beta^2 c {t^{ob.}}^2} \left[
  E_1(z_{\min}) - E_1(z_{\max}) \right].  
\ea
\label{eq:F1} 
\eeq
At a given observed time $t^{ob.}$, the integration boundaries are
$r_{\max} = \beta c t^{ob.}/u_{\min} = \Gamma^2 \beta c
t^{ob.}/(1+\beta)$, and $r_{\min} = \beta c t^{ob.}/(1+\beta)$. In
evaluating the integral in the second line, I use $z \equiv r_0/r$,
which lead to $z_{\max} = (1+\beta) r_0 /(\beta c t^{ob.})$ and
$z_{\min} = (1-\beta) r_0 /(\beta c t^{ob.})$. These can be written
with the use of normalized time, $t_N \equiv r_0(1-\beta)/c$ as
$z_{\max} = [(1+\beta)/(1-\beta)] (t_N/\beta t^{ob.})$ and $z_{\min} =
t_N/\beta t^{ob.}$. In the final formula, $E_1(z) \equiv
\int_z^{\infty} e^{-t} t^{-1} dt$ is the exponential integral.

At late times, $t^{ob.} \gg t_N$, the
difference of the two exponential integrals can be written as
$E_1(z_{\min}) - E_1(z_{\max}) \approx \log(z_{\max}/z_{\min})$, which
enables to write the temporal decay of the observed flux as
\beq
F(t^{ob.}) \approx F_0 {r_0 \over 2 \Gamma^2 \beta^2 c {t^{ob.}}^2} 
\log\left( {1+ \beta \over 1-\beta} \right).
\label{eq:F_final} 
\eeq  
I thus find that the thermal flux decays at late times as $F(t^{ob.})
\propto {t^{ob.}}^{-2}$. 

In order to calculate the temporal change in the observed temperature,
the power law index of the photons comoving energy decay with radius,
resulting from the misalignment of the velocity vectors of the
electrons below the photosphere needs to be specified.  The average
values of the functions $ \langle \omega \rangle$ and $\langle \alpha
\rangle$ (eq. \ref{eq:indx}) are determined with the use of the
probability density function $P(u)$ given in equation \ref{eq:P_u}. By
doing so, I assume that the conditions that led to the validity of
equation \ref{eq:P_u} for the last scattering event (i.e., that before
the scattering the photon propagation direction is parallel to the
flow, and the neglection of the dipole approximation) hold for every
scattering below and close to the photosphere. I further use the fact
that $\Gamma \gg 1$ and the average photon scattering angle below the
photosphere $\langle \theta_{\gamma} \rangle \ll 1$ to approximate $u
= 1- \beta \cos \theta_{\gamma} \simeq (\theta_\gamma^2/2 +
1/2 \Gamma^{2})$, and $\sin \theta_{\gamma} \simeq \theta_\gamma$.
Using these approximations in equation \ref{eq:indx}, one obtains
\beq
\ba{lcl}
\langle \omega \rangle & \simeq & 
{\bigg \langle} {\beta \theta_\gamma^2 \over {2 R_d \over 2 \pi} u^2}
 {\bigg \rangle} \simeq
 {\bigg \langle} { 2 \beta \over {2 R_d \over 2 \pi} u} - 
{ \beta \over {2 R_d \over 2 \pi} \Gamma^2 u^2}
 {\bigg \rangle}  ;  \smallskip \nonumber \\
\langle \alpha \rangle & \simeq & 
{\bigg \langle} {1 \over {2 R_d \over 2 \pi} u } {\bigg \rangle}.
\ea
\label{eq:indx2}
\eeq

The mean of the functions $\langle u^{-1} \rangle$,  $\langle u^{-2}
\rangle$ is calculated using equation \ref{eq:P_u},
\beq
\ba{lcl}
\langle u^{-1} \rangle & = & \int_{1-\beta}^{1+\beta} {P(u) \over u} du =
\Gamma^2 ; \smallskip \nonumber \\
\langle u^{-2} \rangle & = & {\Gamma^4 \over 3} ( 3 + \beta^2) \simeq
{4 \over 3} \Gamma^4.
\ea
\label{eq:mean_a_w}
\eeq
Using equation \ref{eq:mean_a_w}, it is readily found that 
$\langle \omega \rangle \simeq (1/3) \beta \Gamma^2 / (R_d/2\pi)$, 
and $\langle \alpha \rangle \simeq (1/2) \Gamma^2 / (R_d/2\pi)$. Using
these results in equation \ref{eq:eps_r} leads to the conclusion that
the comoving photon energy decays with radius as 
\beq
\varepsilon' (r) \propto r^{-{2 \beta \over 3}} \simeq r^{-2/3}.
\label{eq:eps_rf}
\eeq

The arguments leading to equation \ref{eq:eps_rf}, in particular the
requirement $\Delta r_i / r_i \ll 1$ are valid below the photosphere
(see \S\ref{sec:eps_prime} , eq. \ref{eq:dr3}). Therefore, at radii
much larger than $r_0$ a deviation from this law is expected.
Comparison with the numerical results (\S\ref{sec:numeric}, figure
\ref{fig5} below) shows that indeed at large radii $r\gg r_0$ the
photon comoving energy becomes $r$~independent. This result can be
understood since above the photosphere the optical depth is smaller
than unity, and, if a photon is being scattered at all then the number
of scattering it undergoes is no more than one or two at most. In
calculating the observed temperature at late times, I thus assume that
the comoving photon energy decreases with radius as
$\varepsilon'(r) \propto r^{-2/3}$ at $r \leq A_{brk} \times r_0$, and
$\varepsilon'(r) \propto r^0$ at larger radii, where $A_{brk} = few$.

The photons spectral distribution is thermal, resulting from
thermalization processes occurring in regions deep inside the flow,
which are characterized by very high optical depth. As the photons
propagate outwards, their energy decreases, however the thermal
spectrum is unchanged (see further discussion in
\S\ref{sec:summary} below).  At any given instance, an observer
sees photons emitted from a range of radii and angles. Therefore, even
if the comoving energy spectrum of the photons is thermal (black
body), the observed spectrum deviates from black body, and is a grey
body. Nonetheless, being a convolution of black body spectra, the
observed spectrum is not expected to deviate much from black body
spectra.  The effective temperature of the observed spectra is
\beq
T^{ob.}(t^{ob.}) =  \frac{\int_{r_{\min}}^{r_{\max}} P(r)dr
  \int_{u_{\min}}^{u_{\max}} P(u) du T^{ob.}(r,u) \delta \left(
    t^{ob.} = {r u \over \beta c} \right)}
{\int_{r_{\min}}^{r_{\max}} P(r)dr
  \int_{u_{\min}}^{u_{\max}} P(u) du \delta \left(
    t^{ob.} = {r u \over \beta c} \right)},
\label{eq:T1}
\eeq
where $T^{ob.}(r,u) = T'(r) \D = T'(r)/\Gamma u$, and 
\beq
T'(r) = \left\{ 
\ba{ll}
T'_0 \left( {r \over {\bar r}}\right)^{-2/3} & r \leq A_{brk} r_0, \nonumber
\\
 T'_0 \left( {A_{brk} r_0  \over {\bar r}}\right)^{-2/3} & r > A_{brk} r_0. 
\ea\right.
\label{eq:T_prime}
\eeq

In evaluating the numerator in equation \ref{eq:T1}, one needs to
discriminate between two cases. At early times, $t^{ob.}< A_{brk}
t_N/\beta$, $r_{\max} < A_{brk} r_0$, and therefore the break in the comoving
temperature occurs at radius which is outside the integration boundaries. 
At these times, equation \ref{eq:T1} becomes
\beq
\ba{lcl}
T^{ob.}(t^{ob.}; t^{ob.} \leq A_{brk} t_N) & = &  
{T'_0 \over \Gamma} {{\bar r}^{2/3} \over \beta c t^{ob.}} 
\frac{\int _{r_{\min}}^{r_{\max}} {r_0 \over r^{2/3}} e^{-(r_0/r)} dr}
{\int _{r_{\min}}^{r_{\max}} {r_0 \over r} e^{-(r_0/r)} dr} \nonumber \\
& = & {T'_0 \over \Gamma} {{\bar r}^{2/3} r_0^{1/3} \over \beta c
  t^{ob.}} \Gamma\left(-{1 \over3}\right) \smallskip  \nonumber \\
& & \times 
\frac{ \left[P\left(-{1 \over
        3},z_{\max}\right)  -
    P\left(-{ 1 \over 3},z_{\min}\right)\right]}
{ E_1 (z_{\min}) - E_1(z_{\max}) }. 
\ea
\label{eq:T_final1}
\eeq
Here, $P(a,z) \equiv [1/\Gamma(a)] \int_0^z e^{-t} t^{a-1} dt$ is
incomplete Gamma function\footnote{Note that $\Gamma(a)$ is Gamma
  function with the parameter $a$, not to be confused with the Lorentz
  factor, $\Gamma$.}. At early enough times, $t^{ob.} \ll
t_N$, $z_{\max} \gg z_{\min} \gg 1$. Equation \ref{eq:T_final1} can be
put in a simpler form by expanding the exponential integrals $E_1(z)
\simeq e^{-z}/z$ and the incomplete Gamma function, $\Gamma(-1/3) P
(-1/3,z) \simeq -e^{-z}/z^{4/3}$. With these approximations, equation
\ref{eq:T_final1} becomes
\beq
\ba{lcl}
T^{ob.}(t^{ob.}; t^{ob.} \leq A_{brk} t_N) & \simeq &  
{T'_0 \over \Gamma} {{\bar r}^{2/3} r_0^{1/3}
  \over \beta c t^{ob.} z_{\min}^{1/3}} \smallskip \nonumber \\
& = & { T'_0 {\bar r}^{2/3} (1 +
  \beta)^{1/3} \over \Gamma^{1/3} (\beta c t^{ob.})^{2/3}}.
\ea
\label{eq:T_final1b}
\eeq   

At later observed times $t^{ob.} > A_{brk} t_N/\beta$, the break in
the comoving temperature occurs at radius which is inside the
integration boundaries. Splitting the integral over $r$ in the
numerator of equation \ref{eq:T1} into two, one obtains
\beq
\ba{lcl}
T^{ob.}(t^{ob.}; t^{ob.}  \geq  A_{brk} t_N) & = & 
{T'_0 \over \Gamma} {{\bar r}^{2/3} r_0^{1/3} \over \beta c
  t^{ob.}} \nonumber \\
& & \smallskip \times
\frac{ I_1 + I_2}
{ E_1 (z_{\min}) - E_1(z_{\max}) },
\ea
\label{eq:T_final2a}
\eeq
where
\beq
\ba{lcl}
I_1 & = & \Gamma\left(-{1 \over3}\right) \left[P\left(-{1 \over
        3},z_{\max}\right)  -
    P\left(-{ 1 \over 3},{1 \over A_{brk}} \right)\right], \nonumber \\
I_2 & = & 
A_{brk}^{-2/3} \left[ {e^{-z_{\min}} \over z_{\min}} - A_{brk}
  e^{-1/A_{brk}} \right. \nonumber \\ 
& & \qquad \qquad  + E_1(A_{brk}^{-1}) -E_1(z_{\min})\Big{]}.
\ea
\label{eq:T_final2b}
\eeq
  
Unfortunately, for the relevant time scale $t^{ob.}/t_N \lesssim
10^4$, (see \S\ref{sec:numeric}) there is no simpler analytic
approximation to equation \ref{eq:T_final2a}.

\section{Numerical calculation of the flux and temperature decay at
  late times}
\label{sec:numeric}

The analytical calculations presented above were checked with a
numerical code. The code is a Monte-Carlo simulation, based on earlier
code developed for the study of photon propagation in relativistically
expanding plasma \citep{PW04, PMR06b}. I give below a short
description of the numerical code, before presenting the numerical
results and a comparison to the analytical approximations developed
above.

\subsection{The numerical model}
\label{sec:code}

I consider a three-dimensional plasma wind expanding from an
initial radius $r_i$ that fills the entire volume $r> r_i$. Following
the standard dynamics of GRB outflow \citep[e.g.,][]{MR00}, the plasma
assumed to accelerate up to the saturation radius $r_s = \Gamma r_i$,
above which it expands at constant Lorentz factor,
$\Gamma$. The plasma is assumed to be ejected at a time independent
rate, $\dot{M}= L/\Gamma c^2$, where $L$ is the (observed)
luminosity. Therefore, the plasma comoving density decreases with
radius above $r_s$ as $n'(r) = \dot{M}/(4 m_p \Gamma \beta c r^2)$. Since
below $r_s$, $\Gamma(r) \propto r$, considering adiabatic energy
losses the plasma comoving temperature at $r \geq r_s$ is given by 
\beq
\ba{lcl}
T'_{el}(r) & = & {k_B \over \Gamma m_e c^2} \left( {L \over 4 \pi r_i^2 c a}
\right)^{1/4} \left( {r \over r_s} \right)^{-2/3} \smallskip \nonumber \\
& = & 7.3 \times 10^{-3}
\, \Li^{1/4} r_{i,8}^{1/6} \Gi^{-1/3} r_{10}^{-2/3}.
\label{eq:T_el_num}
\ea
\eeq   
Here, $T'_{el}$ is given in normalized units of $m_e c^2$ and
the convention $Q=10^xQ_x$ is adopted in cgs units. 

Photons are injected into the plasma in a random position on the
surface of a sphere at radius $r_{inj} = R_d/(2\pi \Gamma^2 d)$. The
depth $d$ is taken as $d=20$ in order to ensure that the probability
of a photon to escape without being scattered is smaller than
$\exp(-20)$, i.e., negligible\footnote{In fact, from equations
  \ref{eq:exp} and \ref{eq:mean_a_w} it is found that the average
  number of scattering prior to photon escape is $\approx 2 d$. This
  result was confirmed numerically.}. The initial photon propagation
direction is random, and its (local) comoving energy at the injection
radius is equal to the plasma comoving temperature at this radius.

Given the position of the scattering event and the photon 4-vector,
the position of the next scattering event is calculated as follows.
The code transforms the scattering position into the cylindrical
coordinate system presented in \S\ref{sec:r_ph}, in which the
scattering position is  ($r_{\min}, z_{\min}$). Using
$\theta=\tan^{-1}(r_{\min}/z_{\min})$, the code calculates the optical
depth for the photon to escape, using equation \ref{eq:tau1}. It then
draws an optical depth $\Delta \tau$ from a logarithmic distribution,
which represents the optical depth traveled by the photon until the
next scattering event. If $\Delta \tau$ is larger than the
optical depth to escape, the photon assumes to escape, and the time
difference between this photon and a hypothetical photon that
propagated parallel to the last propagation direction of the photon
and was not scattered is calculated.

If $\Delta \tau$ is smaller than the optical depth to escape, the next
scattering position is calculated along the photon propagation
direction, i.e., it occurs in position ($r_{\min}, z_{\max}$).
$z_{\max}$ is calculated such that the difference in optical depths
between the initial scattering point and the next scattering point is
equal to $\Delta \tau$. The scattering position is then transformed
back to the standard Cartesian coordinates.

Given the scattering position, the electrons temperature is drawn from
a Maxwellian distribution with temperature given by equation
\ref{eq:T_el_num}. The photon 4-vector is Lorentz transformed twice:
first into the (local) frame of the bulk motion of the flow, which
assumed to move at constant Lorentz factor $\Gamma$ in the radial
direction. Second, transformation into the electrons rest frame (the
electron assumed to move randomly within the bulk motion frame). The
photon interact with the electrons via Compton scattering. The full
Klein-Nishina cross section is considered in the interaction, in which
the scattered angles of the outgoing photon are drawn. The outgoing
photon energy and its new propagation direction are calculated in the
electrons' rest frame. The program then Lorentz transforms the photon
4-vector back into the lab (Cartesian) frame, and repeats the
calculation until the photon escapes.

\subsection{Numerical results}
\label{sec:numerical_results}

\begin{figure}
\plotone{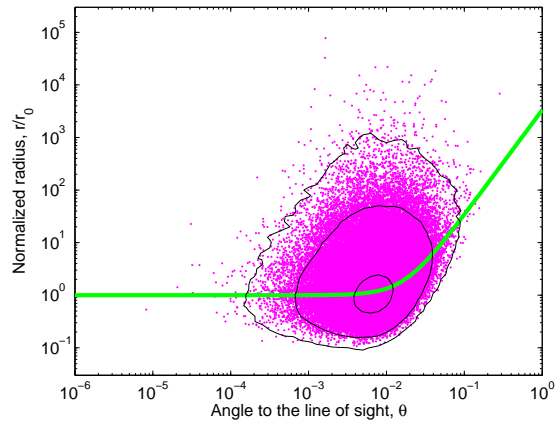}
\caption{ Position of the last scattering event point in $\theta$-$r$
  plane for $10^5$ events. The solid (green) line is the photospheric
  radius, calculated in equation \ref{eq:r_ph}. Clearly, the last
  scattering events take place in a range of radii and angles. The
  photospheric radius gives a first order approximation to the
  position of these events. The contour lines are added to the plot in
  order to indicate the density of the emerging photons radii and
  angles.}
\label{fig3}
\end{figure}

The position of the last scattering events for $N=10^5$ simulated
photon propagation inside the expanding plasma is presented in figure
\ref{fig3}. The last scattering event points are shown in the
$r-\theta$ plane, on top of the photospheric radius calculated in
equations \ref{eq:tau1} and \ref{eq:r_ph}. In preparing the plot,
parameters characterizing GRBs (see eq. \ref{eq:T_el_num}) were
taken. In the figure, the last scattering event radius is normalized
to $r_0 = R_d/2\pi\Gamma^2$. Therefore, results obtained for arbitrary
values of the free model parameters ($L$ and $\Gamma$) that
characterize astrophysical transients other than GRBs, such as AGNs or
microquasars are similar to the ones presented. Clearly, the
photospheric radius calculated in equation \ref{eq:r_ph} gives a first
order approximation of the last scattering events radii and
angles. However, it is obvious from the figure that photons decouple
from the plasma at a range of radii and angles, necessitating the use
of the probability density functions calculated in
\S\ref{sec:analytic}.

\begin{figure}
\plotone{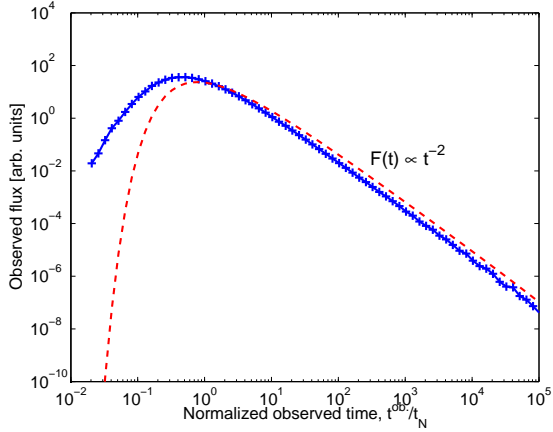}
\caption{Observed thermal flux as a function of the observed time. The
  solid (blue) line is the numerical simulation result, and the dash
  (red) line is the analytic approximation in equation \ref{eq:F1}. Time
  is given in units of normalized time, $t_N$. For $t/t_N \gtrsim 1$,
  the thermal flux decays as $F(t^{ob.}) \propto {t^{ob.}}^{-2} $ }
\label{fig4}
\end{figure}

The normalized observed flux is presented in figure \ref{fig4},
together with the analytical result in equation \ref{eq:F1}. In
producing this figure, the photons are assumed to be injected as a
$\delta$-function in time. As the photons propagate outwards, the
program traces the time in which every scattering event occurs. Once the
photon propagation direction after the last scattering event is known,
calculation of the observed time is done by calculating the lag of the
particular photon with respect to a hypothetical (``trigger'') photon
that propagated in a direction parallel to the final propagation
direction of the photon, and did not undergo any scatterings.

Physical transient sources emit during a finite time duration.  Since
the photon injection function is assumed here to be a $\delta$
function in time, it is clear that the results presented are valid for
late time emission only, after the central engine that produced the
photon emission had decayed. The early time rise in the flux is thus
expected to deviate from the results presented here, and depend on the
properties of the emission mechanism in physical transients. The late
time power law decay, $F(t^{ob.}) \propto {t^{ob.}}^{-2}$ is a prediction of
the model. The time scale in figure \ref{fig4} is presented in
normalized units of $t_N$, and thus is useful for any transient
sources.

For parameters characterizing GRBs, $t_N \simeq 10^{-2} \, \Li
\Gi^{-5}$~s. Since, as calculated above (see \S\ref{sec:time},
eq. \ref{eq:t_max}), thermal emission from GRBs is expected to last
tens of seconds, the relevant time scale is $t/t_N \approx 10^4$. It
is shown in figure \ref{fig4} that on this time scale the
approximation in equations \ref{eq:F1}, \ref{eq:F_final} are in
excellent agreement with the numerical results. This fact confirms the
validity of the approximations introduced in the analytical
calculations in \S\ref{sec:analytic}.

The numerical results show that at very early times, $t/t_N <
10^{-1}$, the analytical result presented in equation \ref{eq:F1} does
not predict the high flux expected. This indicates the limitations of
the model, in particular the assumption that the observed time depends
only on two parameters, the last scattering event radius and angle.
As the photons diffuse below the photosphere, inevitably there is a
small spreading in their arrival times to radius $r$, which is not
considered in the analytical calculations. This discrepancy, however,
is limited to very short time scales, in which, as discussed above,
the actual nature of the inner engine activity determines the observed
flux.

\begin{figure}
\plotone{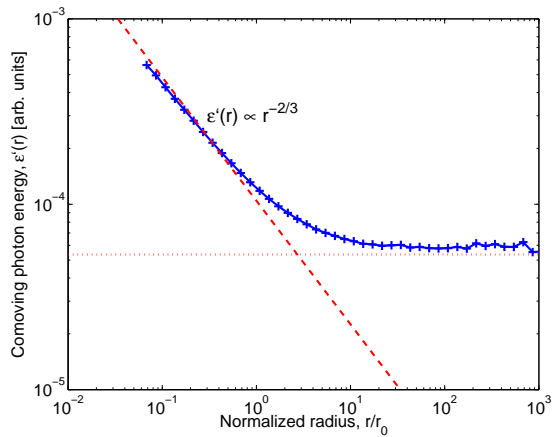}
\caption{Photon (local) comoving energy as a function of radius, at
  the last scattering event radius. Numerical results are presented by the
  solid (blue) line. The results indicate that for $r/r_0 < 1$,
  $\varepsilon'(r) \propto r^{-2/3}$, in accordance with the results of
  equation \ref{eq:eps_rf}.  At larger radii, the arguments leading to
  the result in equation \ref{eq:eps_rf} no longer hold, and the photon
  (local) comoving energy becomes $r$-independent.  }
\label{fig5}
\end{figure}

The (local) comoving photon energy at the last scattering event radius
is presented in figure \ref{fig5}. It is shown that indeed for
$r<r_0$, $\varepsilon'(r) \propto r^{-2/3}$, as calculated in equation
\ref{eq:eps_rf}. This justifies the approximations that led to that
equation. At larger radii, the approximations that led to equation
\ref{eq:eps_rf} no longer hold, as the number of scattering a photon
undergo above this radius is no more than a few. As a result, the photon
comoving energy becomes $r$-independent.

The decay law index of the electrons comoving temperature in equation
\ref{eq:T_el_num}, $2/3$, is similar to the decay law index of
the photon temperature calculated in \S\ref{sec:eps_prime}. In order
to check that the mechanism described in \S\ref{sec:eps_prime} is
indeed independent on the decay law index of the electrons comoving
temperature, additional runs with decay law index larger than $2/3$
were performed. The results obtained in these runs were similar to the
results presented in figure \ref{fig5}.

\begin{figure}
\plotone{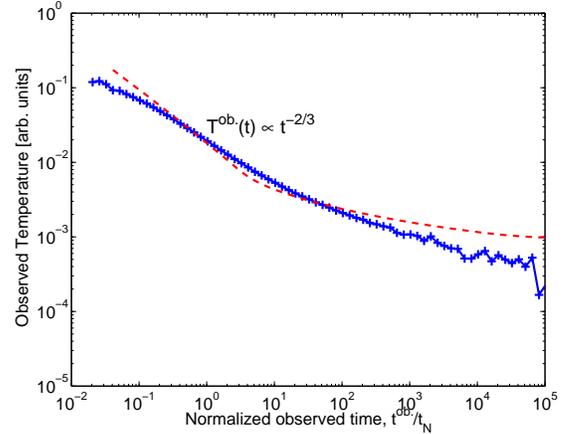}
\caption{Observed temperature as a function of the observed time.  The
  solid (blue) line is the numerical simulation results, and the dash
  (red) line is the analytic approximation in equations
  \ref{eq:T_final1} and \ref{eq:T_final2a}, with $A_{brk} = 3$. The
  numerical results are slightly smoother than the analytical
  approximation. The temperature decays as a power law in time, with
  power law index $\beta \simeq 2/3$ at $t/t_N \lesssim 10$, and
  $\beta \approx 1/2$ at later times.}
\label{fig6}
\end{figure}

The observed temperature as a function of time is presented in figure
\ref{fig6}. The numerical results are shown by the blue solid line,
and the analytical approximation calculated in equations
\ref{eq:T_final1} and \ref{eq:T_final2a} are presented by the red
dashed line. In preparing the plots, $A_{brk} = 3$ was taken, in
accordance to the numerical results of the comoving energy decay at
large radii (see figure \ref{fig5}).  Clearly, the analytical formula
gives a good approximation to the numerical results, although the
numerical results show that the power law decay of the temperature is
somewhat steeper than the analytical approximation at late times:
$T^{ob.}(t^{ob.})  \propto {t^{ob.}}^{-\beta}$, with $\beta \simeq
2/3$ at $t/t_N \lesssim 30$ becoming $\beta \approx 1/2$ at later
times. The model presented here thus predicts a late time power law
decay of the observed temperature, with power law index $\beta$ that
slightly decreases over a time scale of $\sim 100 t_N$ from $\beta
\simeq 2/3$ to $\beta \approx 1/2$. For parameters characterizing
GRBs, this time scale corresponds to few seconds.

\section{Summary and discussion}
\label{sec:summary}

In this paper, I addressed the question of late time thermal emission
from optically thick, relativistically expanding plasma winds. I first
showed in \S\ref{sec:r_ph} that the photospheric radius depends on the
angle to the line of sight, $\theta$ in a non-trivial way: for $\theta
< \Gamma^{-1}$, $r_{ph}$ is $\theta$-independent, while for $\theta >
\Gamma^{-1}$, $r_{ph}(\theta) \propto \theta^2$ (eq. \ref{eq:r_ph}).
I used this result in \S\ref{sec:time} to show that for parameters
characterizing emission from GRBs, thermal emission can be seen up to
tens of seconds following the decay of the inner engine
(eq. \ref{eq:t_max}).  In \S\ref{sec:eps_prime} I showed that photons
lose their energy due to repeated Compton scattering below the
photosphere. The mechanism responsible for this energy loss is based
on the geometrical effect of the misalignment between the velocity
vectors of the electrons in the expanding plasma jet. This mechanism
is therefore unrelated to other mechanisms discussed so far in the
literature (e.g., adiabatic expansion, or the mechanism leading to
Kompaneets equation). I showed that as a result of this mechanism, the
(local) comoving temperature of the photons decreases below the
photosphere as $\varepsilon'(r) \propto r^{-2/3}$
(eq. \ref{eq:eps_rf}). I introduced in \S\ref{sec:analytic} the
probability density function $P(r,\theta)=P(r)\times P(\theta)$ that
extend the definition of a photosphere to include the actual positions
and angles in space from which thermal photons decouple from the
plasma. Using these functions, I calculated the temporal decay of the
observed flux (eqs. \ref{eq:F1}, \ref{eq:F_final}) and temperature
(eqs.  \ref{eq:T_final1}, \ref{eq:T_final2a}) of the thermal
emission, and showed that both decay as a power law in time,
following the decay of the inner engine that produces the thermal
photons.  The flux decays as $F(t^{ob.}) \propto {t^{ob.}}^{-\alpha}$
with $\alpha = 2$, and the temperature decays as $T(t^{ob.}) \propto
{t^{ob.}}^{-\beta}$ with $\beta = 2/3$ at early times which later
changes to $\beta \simeq 1/2$. The analytical results were confirmed
with the results of the numerical simulation presented in
\S\ref{sec:numeric}.

The results presented here can account for the recent observations of
thermal emission that accompanies long duration GRBs. As was shown by
\citet{Ryde04, Ryde05}, after $\sim 1-3$~s the temperature of the
thermal component decreases as a power law in time, $T^{ob.} \propto
t^{-\alpha}$, with power law index $\alpha \simeq 0.6 - 1.1$. An
additional analysis (F. Ryde \& A.Pe'er 2008, in preparation) shows
that after a short rise, the flux of the black body component of these
bursts also decreases with time as $F_{BB}^{ob.} \propto t^{-\beta}$,
with power law index $\beta \approx 2.0 - 2.5$. These results are thus
naturally reproduced by the model presented here.

A key consequence of the model presented here is that thermal emission
at early times (before the observed break) is dominated by thermal
photons originating from the photosphere on the line of
sight. Therefore, observation of the thermal emission at early times,
when the inner engine is still active, gives a direct measurement of the
temperature and flux of photons emitted from the photospheric radius
on the line of sight, $r_0 \equiv r_{ph}(\theta=0)$. This is the
innermost radius from which information can reach the observer.

The interpretation presented here has a direct implication in the
study of relativistic outflows.  For GRBs with known redshift, early
time (before the break) observation of the temperature and flux of the
thermal component enabled direct determination of two of the least
restricted parameters of the fireball model: the bulk motion Lorentz
factor, $\Gamma$ and the radius at the base of the flow
\citep{PRWMR07}. Being based on thermal emission only, the method
presented in this paper is insensitive to many of the inherent
uncertainties in former methods of determining the values of these
parameters. Future measurements with the upcoming {\it GLAST}
satellite will enable to increase the sample of GRBs with known
redshift from which thermal emission component is identified, to
further test the model presented here and to gain statistics on the
values of the fireball model parameters.

In addition to the prompt emission phase in GRBs, thermal activity may
occur as part of the flaring activity observed in the early afterglow
phase of many GRBs \citep{Burrows05, Falcone07}. The exact nature of
these flares is currently not yet clear. As it is plausible that the
flares result from renewed emission from the inner core, a renewed
thermal emission may occur. Analyzing this emission in a method
similar to the one described here and by \citet{PRWMR07}, may thus
provide information on the flow parameters during the late time
flaring activity.

The relevance of the results obtained here is not limited only to
emission from GRBs, but also to emission from any transient phenomenon
characterized by relativistic outflow, such as AGNs and
microquasars. Provided there is a source of photons deep inside the
flow, following the decay of this source the decay laws of the thermal
flux and temperature derived above hold for any such object. An
important point here, is that the nature of the mechanism that produces
the radiation is of no importance, as long as it occurs deep inside
the flow so that the photons thermalize before they escape.

In this work, I assumed that the electrons are cold (in the comoving
frame), and that the electrons and photons interact only via Compton
scattering. If this is not the case, due, e.g., to some dissipation
mechanism that produces energetic electrons at different regions of the
flow, than Compton scattering with energetic electrons will lead to
modification of the thermal spectrum. This case was extensively
studied by \citet{RM05, PMR05, PMR06}. As was shown in these works,
the thermal photons in this case serve as seed photons to Compton
scattering that produces high energy, non-thermal spectrum. However,
if the optical depth in which the energetic electrons are introduced
into the flow is smaller than $\sim$ unity, then the thermal component
can be separated from the non-thermal one \citep{PMR06}. Note that
this is exactly the case in the internal collision model of GRBs:
internal shocks can only occur at radii larger than the spreading
radius, $r_{spread} = \Gamma r_S = 10^{12} \, r_{i,8} \Gi^2$~cm, which
is similar to $r_0 = 6 \times 10^{12} \, \Li \Gi^{-3}$~cm. Thus, the
optical depth in the region where internal shocks can occur is not
expected to exceed a few. I can thus conclude that thermal emission is
expected to be observed in GRBs under the assumptions of the internal
collisions model.

The late time thermal emission predicted here essentially arises from
emission off the line of sight. It is thus similar in nature to the
high latitude emission discussed in the literature, in the context of
GRB afterglow emission \citep{FMN96,WL99,KP00}. All these works,
however, treated the optically thin case, which is relevant for the
afterglow emission phase from GRBs. The work presented here differs
by treating thermal emission from optically thick plasmas,
characterizing the very early stages of emission from GRBs.

One of the key findings in this work is the new mechanism in which
photons lose their energy below the photosphere. This mechanism
differs than other mechanisms discussed in the literature so far
for radiative cooling below the photosphere. The result obtained,
$\varepsilon'(r) \propto r^{-2/3}$ holds  for relativistic jets
characterized by constant ($r$-independent) jet opening angle. For
jets in which the jet opening angle is $r$-dependent, a different
power law decay in the photon energy is expected. 

In the calculation of the photon energy loss presented in
\S\ref{sec:eps_prime}, I neglected the electrons temperature. As the
photons propagate downstream, their comoving temperature cannot be
lower than the comoving temperature of the electrons. The
electrons temperature decreases due to adiabatic expansion, which
result in a decay of the electrons temperature as a power law in the
comoving plasma volume, $T'_{el} \propto {V'}^{-1/3}$ (for relativistic
electrons). Adopting the fireball model of GRBs \citep[for review,
see, e.g.][]{Mes06}, above the saturation and below the spreading
radii of the fireball, the comoving volume is $V' \propto r^{2}$,
resulting in a decrease of the comoving electrons temperature as
$T'_{el}(r) \propto r^{-2/3}$. Above the spreading radius,
the comoving volume increases as $V' \propto r^3$, which implies
$T'_{el}(r) \propto r^{-1}$. In any of these regimes, the electrons
temperature decreases with the radius at least as fast as the photon
temperature.

The mechanism presented here for photon energy loss has some
resemblance to adiabatic energy losses, as the photon temperature
is converted into work done on the electrons. However, it
is a different mechanism having a different origin. Adiabatic energy
losses occur once the plasma expands, and its volume increases. As
opposed to that, in the scenario considered here, the volume in which
photons interact with the electrons (the volume below the photosphere)
does not expand, since for constant flow parameters (${\dot{M}},
\Gamma$) the photospheric radius is time independent. In addition, as
discussed above, the decay law of the photon energy is independent on
the electrons temperature (as long as the electrons comoving
temperature is not higher than the photon comoving temperature). The
fact that between the saturation radius and the spreading radius in
GRBs the decay law of the photons and electrons comoving temperature
is similar, is thus a coincidence.

\acknowledgments 

I would like to thank Felix Ryde for many useful discussions that
initiated and accompanied this work. I would also like to thank Ralph
A.M.J. Wijers, Peter M\'esz\'aros, Mario Livio, Andy Fruchter, Julian
Krolik and Ron Allen for many useful discussions and comments. I wish
to express my warmest gratitude to Martin J. Rees, without him this
work could not have been made possible. This research was supported by
the Riccardo Giacconi Fellowship award of the Space Telescope Science
Institute.

\end{document}